\documentclass[%
 reprint,
groupedaddress,
 amsmath,amssymb,
 aps,
 pra,
]{revtex4-2}
\usepackage{float}
\usepackage{color}
\usepackage{graphicx}% Include figure files
\usepackage{dcolumn}% Align table columns on decimal point
\usepackage{bm}% bold math
\begin{document}

%\preprint{APS/123-QED}

\title{Broadband Optical Two-Dimensional Coherent Spectroscopy of a Rubidium Atomic Vapor}

\author{Jieli Yan}
 \affiliation{Department of Physics, Florida International University, Miami, FL 33199, USA}

\author{Stephen Revesz}
 \affiliation{Department of Physics, Florida International University, Miami, FL 33199, USA}
 
\author{Danfu Liang}
 \affiliation{Department of Physics, Florida International University, Miami, FL 33199, USA}
 
%\author{...}

\author{Hebin Li}
\email{hebin.li@fiu.edu}
 \affiliation{Department of Physics, Florida International University, Miami, FL 33199, USA}

\date{\today}% It is always \today, today,

\begin{abstract}
Optical two-dimensional coherent spectroscopy (2DCS) has become a powerful tool for studying energy level structure, dynamics, and coupling in many systems including atomic ensembles. Various types of two-dimensional (2D) spectra, including the so-called single-quantum, zero-quantum, and double-quantum 2D spectra, of both D lines (D$_1$ and D$_2$ transitions) of potassium (K) atoms have been reported previously. For rubidium (Rb), a major difference is that the D-lines are about 15 nm apart as opposed to only about 3 nm for K. Simultaneously exciting both D-lines of Rb atoms requires a broader laser bandwidth for the experiment. Here, we report a broadband optical 2DCS experiment in an Rb atomic vapor. A complete set of single-quantum, zero-quantum, and double-quantum 2D spectra including both D-lines of Rb atoms were obtained. The experimental spectra were reproduced by simulated 2D spectra based on the perturbation solutions to the optical Bloch equations. This work in Rb atoms complements previous 2DCS studies of K and Rb with a narrower bandwidth that covers two D-lines of K or only a single D-line of Rb. The broadband excitation enables the capability to perform double-quantum and multi-quantum 2DCS of both D-lines of Rb to study many-body interactions and correlations in comparison with K atoms. 
\end{abstract}

\maketitle

\section{Introduction}
Optical two-dimensional coherent spectroscopy (2DCS) \cite{Cundiff2013,Li2017} is an optical analog to two-dimensional nuclear magnetic resonance (NMR) spectroscopy \cite{Ernstbook}. The idea of implementing 2DCS in the optical regime was proposed by Tanimura and Mukamel in 1993 and was realized experimentally first by using infrared ultrafast pulses \cite{Golonzka2001,Hamm1999a}. Since then, various approaches \cite{Bristow2008,Brixner2004,Chen2008,Cowan2004,Li2013,Nardin2013,Selig2008a,Tekavec2007,Turner2011,Volkov2005} have been developed in the near-infrared and visible range. By unfolding a potentially congested 1D spectrum onto a 2D plane and correlating the dynamics in two different frequency dimensions, optical 2DCS excels in studying optical response of complex systems. The technique has become a powerful spectroscopic tool to study energy level structure, dynamics, and coupling in various systems, such as structural information in proteins \cite{Hamm1999a}, dynamics of hydrogen bond in water \cite{Fecko2003a}, energy transfer processes in photosynthesis \cite{Brixner2005,Engel2007,Collini2010}, many-body interactions and correlations in atomic ensembles \cite{Dai2012a,Gao:16,PhysRevLett.120.233401,Yu2019,Yu2018,Liang2021,Yu:22,PhysRevLett.128.103601}, ultrafast dynamics and couplings of excitons in semiconductor quantum wells \cite{Cundiff2012,Li2006a,PhysRevLett.112.097401,Nardin2014,Singh2013,Stone2009a,Turner2012}, quantum dots \cite{PhysRevB.87.041304,Moody2013b,Moody2013,Moody2013a}, 2D materials \cite{Moody2015,Titze2018}, and perovskites \cite{Jha2018,Monahan2017,Nishida2018,Richter2017,Thouin2018,Titze2019}. 

Potassium (K) and rubidium (Rb) atomic vapors were initially used as model systems to validate optical 2DCS techniques. The energy level structures and other parameters are well characterized and the D-lines can be excited by a typical Ti:sapphire femtosecond oscillator, making K and Rb ideal samples to test optical 2DCS implementations. Several early approaches of optical 2DCS were first demonstrated in Rb \cite{Tian2003,Tekavec2007} and K \cite{Dai2010} vapors. Atomic vapors were also used to develop new advances such as optical 3D coherent spectroscopy \cite{Li2013a} and frequency-comb based optical 2DCS \cite{Lomsadze2017b,Lomsadze2018}, and quantitative analyses such as pulse propagation effects \cite{Li2013b,Spencer2015} and line shape analysis \cite{Namuduri2020} in 2D spectra.

Despite being considered initially as a model system, later optical 2DCS measurements of atomic vapors provided interesting insights into many-body interaction and correlation in atoms. Double-quantum 2DCS revealed two-atom collective resonances due to dipole-dipole interaction in both K \cite{Dai2012a} and Rb \cite{Gao:16,PhysRevLett.120.233401} vapors. The method provides an extremely sensitive and background-free detection to probe interatomic dipole-dipole interaction even in a dilute vapor with a density as low as $4.81\times 10^8$ cm$^{-3}$, corresponding to a mean interatomic separation of 15.8 $\mu$m \cite{Yu2018}. The technique was also extended to multi-quantum 2DCS to probe multi-atom Dicke states with up to eight atoms \cite{Yu2019,Liang2021}. Optical 2DCS can potentially be a useful tool to study many-body physics in cold atoms. For K vapor, a complete set of single-quantum, zero-quantum, and double-quantum 2D spectra involving both D$_1$ and D$_2$ lines have been reported \cite{Dai2010,Dai2012a}. However, previous 2DCS studies of Rb vapor measured only single-quantum 2D spectra \cite{Tian2003,Tekavec2007} of both D-lines or double-quantum 2D spectra of an individual D-line \cite{Gao:16,PhysRevLett.120.233401}. Compared to K, a major difference in Rb is that the D-lines are 15 nm apart, requiring a broader laser bandwidth to cover both D-lines simultaneously. 

Here, we implemented broadband optical 2DCS in an Rb atomic vapor and obtained a complete set of single-quantum, zero-quantum, and double-quantum 2D spectra of both D-lines of the Rb atom. The single-quantum and zero-quantum 2D spectra show the coherent coupling between the two D-line transitions, while the double-quantum 2D spectrum reveals the collective resonances involving both $5P_{1/2}$ and $5P_{3/2}$ states. We also present a theoretical treatment based on the perturbative solution to the optical Bloch equation (OBE). The simulated 2D spectra agree well with the experimental spectra. The rest of the paper is organized as the following. Section \ref{sec1} describes the experimental setup of a collinear 2DCS implementation. Section \ref{sec2} introduces the theoretical approach based on the OBE. Section \ref{sec3} describes the experimental and theoretical details of single-quantum and zero-quantum 2D spectra. Section \ref{sec4} the experimental and theoretical details of double-quantum 2D spectra. Section \ref{sec5} is a conclusion. 

\section{\label{sec1}Experimental setup}
\begin{figure}[tbh]
	\centering
	\includegraphics[width=\columnwidth]{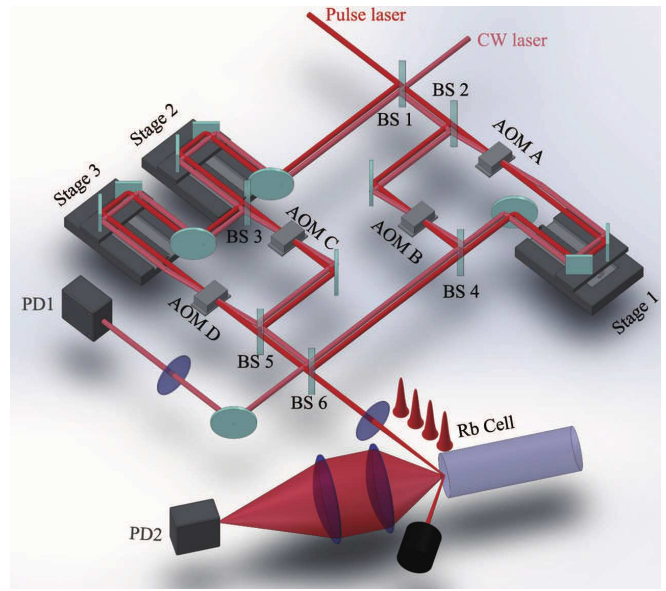}
	\caption{Schematic of the collinear optical 2DCS experimental setup. BS: Beamsplitter. AOM: Acousto-optic modulator. PD: Photo detector.}
	\label{fig:setup}
\end{figure}

The optical 2DCS experiment was performed in a collinear setup based on acousto-optic modulators (AOM) \cite{Nardin2013}. As shown in Fig. \ref{fig:setup}, the primary apparatus consists of an interferometer of two Mach-Zehnder interferometers. This nested interferometer splits an input fs laser pulse into four pulses (A, B, C, and D) and then combines them in one beam with three time delays between the pulses. The delays are controlled by three translation stages. Each pulse goes through an AOM and is phase modulated at a slightly different frequency $\Omega_i$ ($i=$A, B, C, and D). The first-order diffraction from the AOM output is used for the experiment while the zeroth-order output is terminated. Meanwhile, the output of a continuous-wave (CW) laser (an external cavity diode laser) is injected into the interferometer and propagates through the same optical path and the AOMs as the fs laser pulses. The fs laser and CW laser beams are offset such that they can be separated after exiting the interferometer at beamsplitter 6 (BS6). The CW laser signal is detected by a photodetector (PD1). The CW signal carries the AOM modulation frequencies and their beating frequencies, which can be used as the reference for lock-in detection. The CW laser beam also monitors the optical path fluctuations in real time. The fs pulse sequence is incident on the window of an Rb atomic vapor cell which is placed in an oven for temperature control. The resulting fluorescence signal is collected and directed to a photodetector (PD2) by a pair of lenses.    

The fourth-order nonlinear fluorescence signal can be measured by lock-in detection using a proper combination of CW laser beating frequencies as the reference. The nonlinear signal associated with different pulse sequences and phase-matching conditions can be selectively detected by referencing to the proper mixing frequency. For instance, for the rephasing single-quantum signal in the phase-matched direction $\mathbf{k}_S=-\mathbf{k}_A+\mathbf{k}_B+\mathbf{k}_C-\mathbf{k}_D$, where $\mathbf{k}_{A,B,C,S}$ are the wave vectors for pulses A, B, C, and the signal, the reference frequency for lock-in detection should be $\Omega_{S1}=-\Omega_A+\Omega_B+\Omega_C-\Omega_D$. To obtain this reference frequency, the CW laser signal is processed by a digital wave mixer so that the beating frequencies $\Omega_{AB}=\Omega_A-\Omega_B$ and $\Omega_{CD}=\Omega_C-\Omega_D$ are extracted by filtering and subsequently mixed to generate $\Omega_{S1}=\Omega_{CD}-\Omega_{AB}$. The rephasing single-quantum signal can be measured by lock-in detection referencing to $\Omega_{S1}$. Other types of 2DCS signal can be obtained similarly by using different mixing frequencies as the lock-in reference. 
The time delay of each pulse can be controlled independently. There are three time periods, $\tau$ between the first and second pulse, $T$ between the second and third pulse, and $t$ between the third and fourth pulse. During the experiment, the signal is recorded while scanning two or three time delays. For instance, if both $\tau$ and $t$ are scanned, the signal can be represented in the time domain as $S(\tau,t)$ with a fixed $T$. A 2D spectrum $S(\omega_\tau,\omega_t)$ is generated by 2D Fourier-transforming the time-domain signal $S(\tau,t)$ into the frequency domain. Depending on the time ordering of the excitation pulses and which time delays are scanned, the experiment can generate single-quantum, zero-quantum, and double-quantum 2D spectra. The experimental detail and interpretation of these 2D spectra are described in the following sections. 

In the current work, the fs laser pulse is provided by a Ti:sapphire oscillator (Coherent Vitara). The output pulse spectrum has a bandwidth (FWHM) of 67.45 nm with a central wavelength of 810.89 nm. The repetition rate is 80 MHz. The fs laser power at the sample is 18 mW in total. The fs pulses have sufficient bandwidth to simultaneously excite two D-lines of Rb atoms to obtain 2D spectra of both D-lines.

\section{\label{sec2}Theoretical approach}

Using the density matrix formalism, the light-matter interaction in this experiment can be described by the equation of motion \cite{Scully1997}
\begin{equation}
\dot{\rho}_{ij} = -\frac{i}{\hbar} \sum_{k} (H_{ik}\rho_{kj} - \rho_{ik}H_{kj}) - \Gamma_{ij}\rho_{ij}, \label{eq:motion}
\end{equation}
where $\rho_{ij}$ are the density matrix elements. The Hamiltonian has matrix elements $H_{ij}=\hbar\omega_i\delta_{ij}-\mu_{ij}E(t)$, where $\hbar\omega_i$ is the energy of state $|i\rangle$, $E(t)$ is the electric field, $\mu_{ij}$ ($i\neq j$) is the dipole moment of the transition between $|i\rangle$ and $|j\rangle$, and $\delta_{ij}$ is the Kronecker delta function. The relaxation operator $\Gamma$ has matrix elements $\Gamma_{ij}=\frac{1}{2}(\gamma_i+\gamma_j)+\gamma_{ij}^{ph}$, where $\gamma_i$ and $\gamma_j$ are the population decay rates for states $|i\rangle$ and $|j\rangle$, respectively, and $\gamma_{ij}^{ph}$ is the pure coherence dephasing rate ($\gamma_{ij}^{ph}=0$ for $i=j$).   

\begin{figure}[tbh]
	\centering
	\includegraphics[width=\columnwidth]{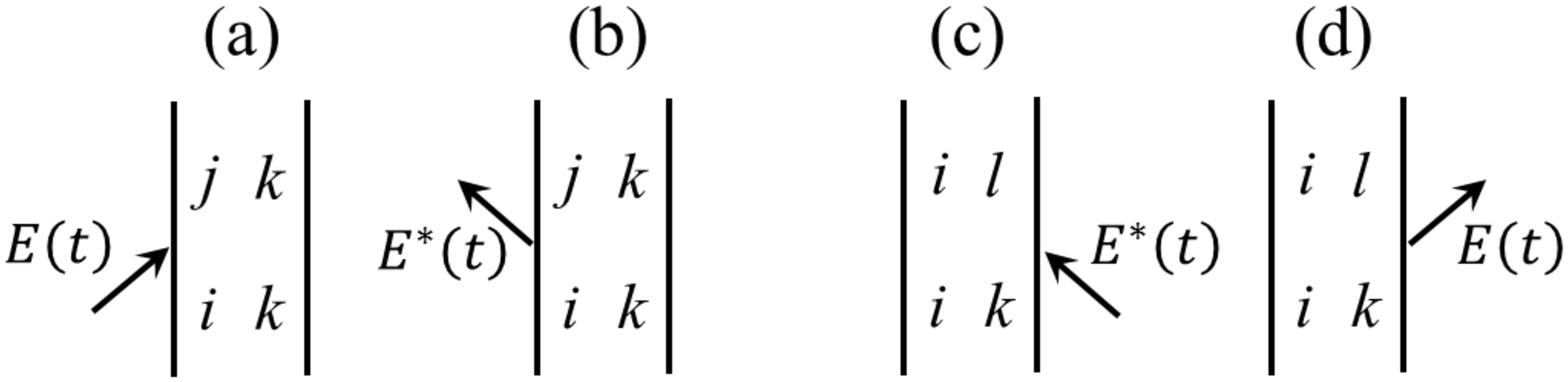}
	\caption{Four different types of vertices in double-sided Feynman diagrams.}
	\label{fig:vertices}
\end{figure}

The equations of motion represented by Eq. (\ref{eq:motion}) include a series of coupled differential equations. To solve these equations perturbatively, we substitute $E$ with $\lambda E$ and expand the density matrix elements as 
\begin{equation}
    \rho_{ij}=\rho_{ij}^{(0)}+\lambda \rho_{ij}^{(1)} + \lambda^2 \rho_{ij}^{(2)}+\lambda^3 \rho_{ij}^{(3)}+ \cdots,
\end{equation}
where $\lambda$ is a constant to track the order. Plug them into Eq. (\ref{eq:motion}) and collect coefficients of $\lambda^n$ ($n=0, 1, 2, \cdots$), we can find the perturbative solutions for each order. In general, the $n$-th order solution of density matrix element is related to the $(n-1)$-th order solution. Under the excitation of a field $\textbf{E}(t)=\hat{E}_n e^{i\mathbf{k}_n\cdot \mathbf{r}-i\omega_n t} +c.c.$, the $(n-1)$-th order density matrix element $\rho_{ik}^{(n-1)}$ evolves to the $n$-th order density matrix element $\rho_{jk}^{(n)}$ or $\rho_{il}^{(n)}$. There are four possibilities that can be calculated by the following integrals
\begin{eqnarray}
\mathrm{(a)}\ \ \ \rho_{jk}^{(n)}&=&\frac{i\mu_{ij}}{2\hbar}e^{i\mathbf{k}_n\cdot \mathbf{r}} \int^t_{-\infty}\hat E_n(t')e^{-i\omega_nt'}\nonumber\\
&&e^{-i\Omega_{jk}(t-t')}\rho^{(n-1)}_{ik}(t')dt' , \label{eq:integral1}\\
\mathrm{(b)}\ \ \ \rho_{jk}^{(n)}&=&\frac{i\mu_{ij}}{2\hbar}e^{-i\mathbf{k}_n\cdot \mathbf{r}}\int^t_{-\infty}\hat E^{*}_n(t')e^{i\omega_nt'} \nonumber\\
&&e^{-i\Omega_{jk}(t-t')}\rho^{(n-1)}_{ik}(t')dt' , \label{eq:integral2}\\
\mathrm{(c)}\ \ \ \rho_{il}^{(n)}&=&-\frac{i\mu_{kl}}{2\hbar}e^{-i\mathbf{k}_n\cdot \mathbf{r}}\int^t_{-\infty}\hat E^{*}_n(t')e^{i\omega_nt'} \nonumber\\
&&e^{-i\Omega_{il}(t-t')}\rho^{(n-1)}_{ik}(t')dt' ,\label{eq:integral3}\\
\mathrm{(d)}\ \ \ \rho_{il}^{(n)}&=&-\frac{i\mu_{kl}}{2\hbar}e^{i\mathbf{k}_n\cdot \mathbf{r}}\int^t_{-\infty}\hat E_n(t')e^{-i\omega_nt'} \nonumber\\
&&e^{-i\Omega_{il}(t-t')}\rho^{(n-1)}_{ik}(t')dt' , \label{eq:integral4}
\end{eqnarray}
where $\Omega_{ij}=\omega_i-\omega_j-i\Gamma_{ij}$. A convenient way to track the time evolution of density matrix elements in the perturbation calculation is to use double-sided Feynman diagram. The interaction with a field is described by the vertex of an arrow with a vertical line. These four integrals can be represented by four vertices, as shown in Fig. \ref{fig:vertices}(a-d), respectively. An arrow represents a field that changes one index of the density matrix element. A photon is absorbed (emitted) if the arrow points towards (away from) the vertical lines. An arrow pointing to the right indicates that the field is $E(t)$, while an arrow pointing to the left means that the field is conjugated, $E^*(t)$. 

A doubled-side Feynman diagram representing an excitation quantum pathway may include multiple orders of these four vertices. The corresponding nonlinear signal can be calculated order by order by using the integrals in Eqs. (\ref{eq:integral1}-\ref{eq:integral4}). In an experiment, the nonlinear signal usually includes contributions from multiple excitation pathways represented by multiple double-sided Feynman diagrams. The contributions from individual diagrams can be calculated separately and added together to obtain the overall nonlinear signal.

\section{\label{sec3}Single-quantum and zero-quantum 2D spectra}

\begin{figure*}[tbh]
	\centering
	\includegraphics[width=\textwidth]{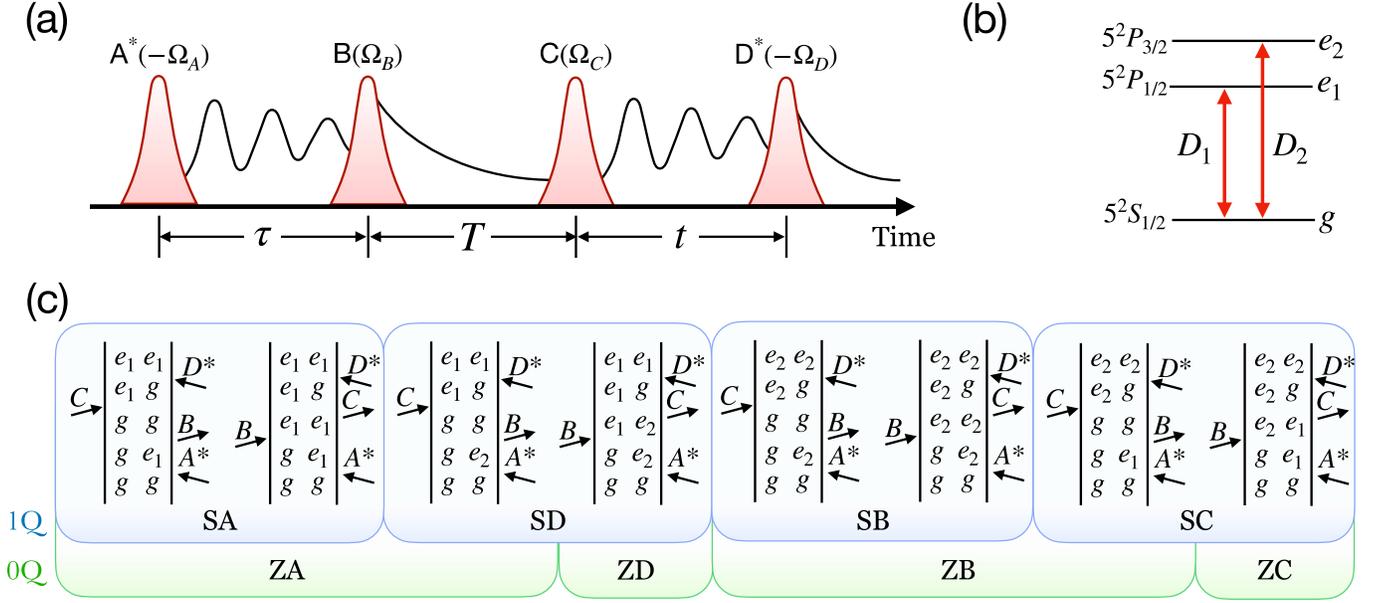}
	\caption{(a) Time ordering of excitation pulses for performing zero-quantum and single-quantum 2DCS. (b) Relevant energy levels of Rb atoms. (c) Double-sided Feynman diagrams representing possible excitation quantum pathways in zero-quantum and single-quantum 2DCS using the pulse sequence in (a).}
	\label{fig:s1schematic}
\end{figure*}

Single-quantum and zero-quantum rephasing 2D spectra were obtained by using the pulse sequence shown in Fig. \ref{fig:s1schematic}(a), in which pulses A and D are considered conjugated. The Rb atom is considered a three-level $V$ system with the relevant energy levels shown in Fig. \ref{fig:s1schematic}(b). A fourth-order nonlinear signal is generated in the sample by the excitation of the four-pulse sequence. Briefly, the first pulse, A, creates coherence between the ground and excited states. The second pulse, B, converts the coherence to a population in either the ground state or the excited states, depending on the relative phase between the first pulses. For a three-level $V$ system, the first two pulses also create a Raman-like coherence between the two excited states. The third pulse, C, converts the population and the ``Raman'' coherence to a third-order coherence between the ground and excited states. The fourth pulse, D, converts the third-order coherence to a population in the excited states which emits fluorescence as the fourth-order nonlinear signal. The dynamics measured during the first time delay $\tau$ reveal the time evolution of the coherence between the ground and excited states and the corresponding coherence dephasing time $T_2$. During the second time delay $T$, the dynamics include the population decay term and the ``Raman'' term. The population term leads to a spectral signal at $\omega_T=0$ which can reveal the population decay time $T_1$, while the ``Raman'' term results in a signal at $\omega_T=\pm \omega_{e_2e_1}$. The process consists of eight specific excitation quantum pathways represented by the double-sided Feynman diagrams shown in Fig. \ref{fig:s1schematic}(c).  

Pulse D can be seen as a local oscillator for heterodyne detection of the third-order coherence within the sample itself. The resulting signal is fluorescence emitted by the fourth-order population. To selectively detect the fourth-order fluorescence signal that is a specific response to the pulse sequence in Fig. \ref{fig:s1schematic}(a), the output of PD2 in Fig. \ref{fig:setup} is demodulated by a lock-in amplifier referenced to the mixing signal $\Omega_{S1}=-\Omega_A+\Omega_B+\Omega_C-\Omega_D$ generated from the CW laser signal. The AOM modulation frequencies used in this experiment are $\Omega_{A}$ = 80 MHz, $\Omega_{B}$ = 80.0173 MHz, $\Omega_{C}$ = 80.104 MHz, and $\Omega_{D}$ = 80.107 MHz so the reference frequency is $\Omega_{S1}=14.3$ kHz.

\begin{figure}[tbh]
	\centering
	\includegraphics[width=0.9\columnwidth]{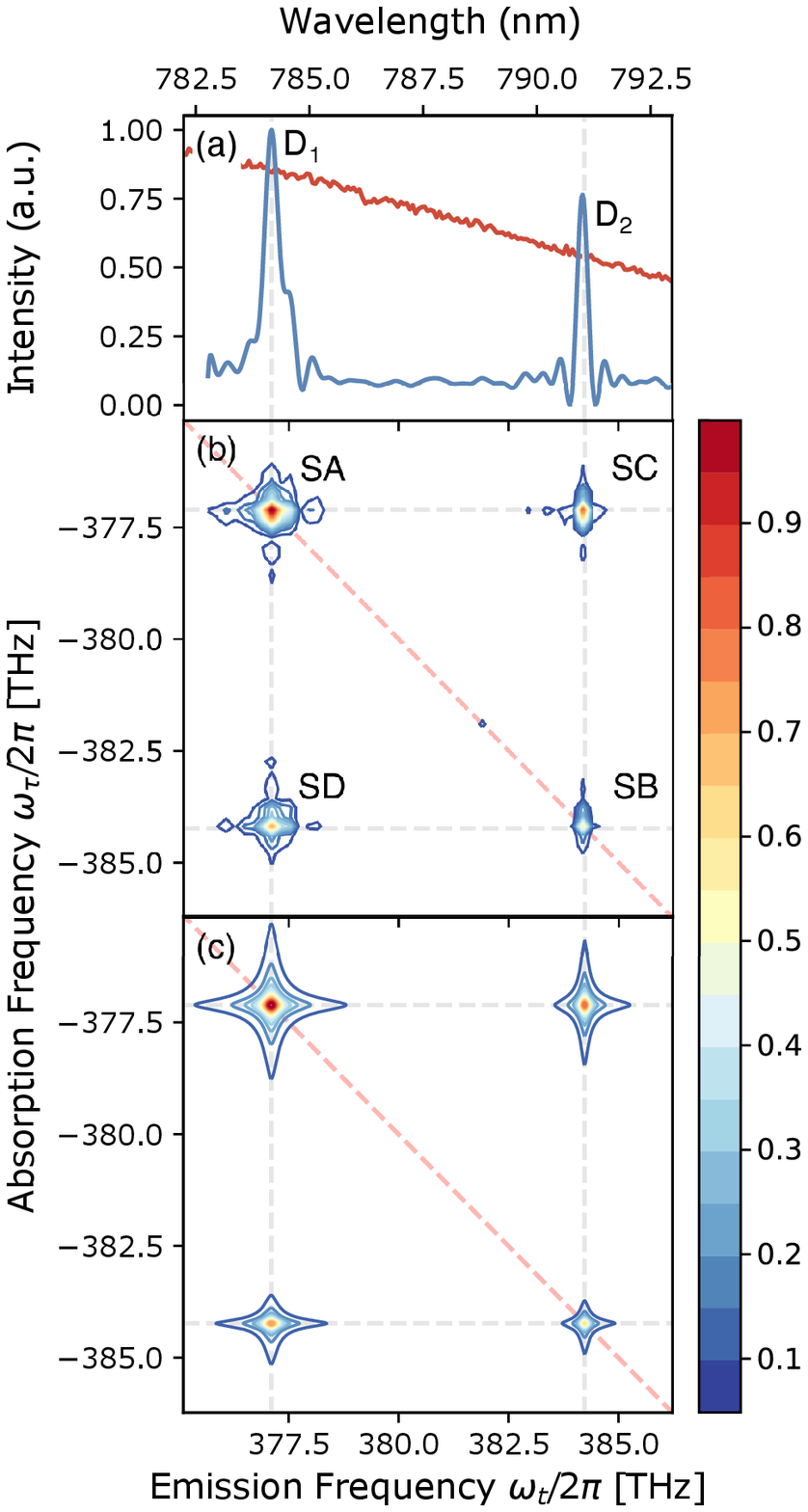}
	\caption{(a) Red curve shows the laser spectrum between 375.107 THZ and 386.23 THz. Note that the laser spectrum is much broader than the spectra range shown here. The projection of the single-quantum spectra onto the emission axis is shown as the blue curve. The D$_{1}$ and D$_{2}$ transitions are shown respectively. (b) Experimental measured single-quantum 2D spectrum of Rubidium vapor at 160 $^\circ$C. (c) Simulated single-quantum 2D spectrum matched to experimental data.}
	\label{fig:s1spectra}
\end{figure}

Single-quantum rephasing 2D spectra can be obtained by scanning time delays $\tau$ and $t$, while fixing the second time delay $T$, and 2D Fourier-transforming the signal into the frequency domain. A typical single-quantum 2D spectrum is shown in Fig. \ref{fig:s1spectra} (a), where the spectral amplitude is plotted with the maximum normalized to 1. The $x$ and $y$ axes are the emission frequency $\omega_t$ and the absorption frequency $\omega_\tau$ corresponding to the time delays $t$ and $\tau$, respectively. The absorption frequency has a negative sign because the excitation pulse A is conjugated. The dotted diagonal line indicates that the absorption and emission frequencies have the same absolute value ($|\omega_\tau|=\omega_t$). Peak SA (SB) on the diagonal line is due to the D$_1$ (D$_2$) transition, absorbing and emitting at the same frequency. In contrast, the off-diagonal peak SC (SD) absorbs at the D$_1$ (D$_2$) frequency and emits at the D$_2$ (D$_1$) frequency, revealing the coupling between the D$_1$ and D$_2$ transitions. Each peak has contributions from two of the double-sided Feynman diagrams in Fig. \ref{fig:s1schematic}(c) which are grouped and labeled accordingly. For peaks SA and SB, the two diagrams represent similar processes. For peaks SC and SD, the two diagrams describe two different dynamics, ground state depletion and ``Raman'' coherence, during the second time delay $T$. The process involving ``Raman'' coherence can be isolated in a zero-quantum 2D spectrum.

\begin{figure}[tbh]
	\centering
	\includegraphics[width=\columnwidth]{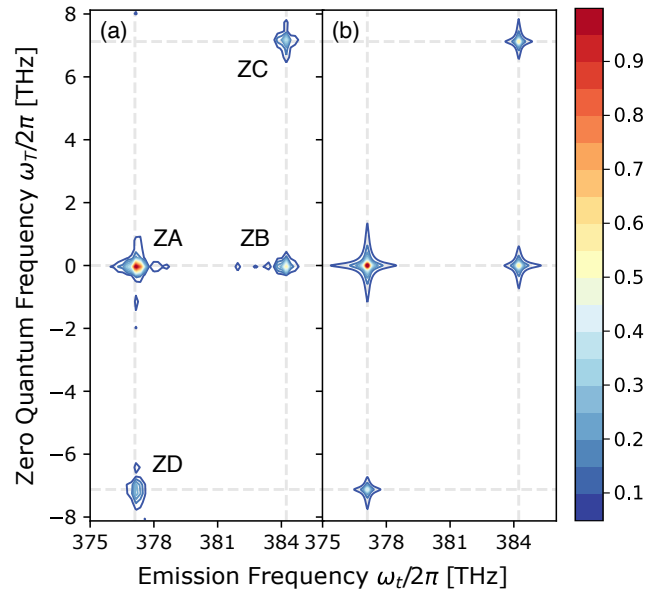}
	\caption{(a) Experimental and (b) simulated zero-quantum 2D spectra of a Rb vapor. The spectral amplitude is plotted with the maximum normalized to 1.}
	\label{fig:s0spectra}
\end{figure}

Zero-quantum 2D spectra can be obtained by scanning time delays $T$ and $t$, while fixing the first time delay $\tau$, and 2D Fourier-transforming the signal into the frequency domain. A typical zero-quantum 2D spectrum is shown in Fig. \ref{fig:s0spectra} (a), where the amplitude is plotted with the maximum normalized to 1. The $x$ axis is the emission frequency $\omega_t$. The $y$ axis is the mixing frequency $\omega_T$ corresponding to the second time delay $T$. There are four peaks in the spectrum. Peaks ZA and ZB are located on the $\omega_T=0$ line. They each are contributed by the three pathways involving a population decay during the second time delay $T$, as labeled in Fig. \ref{fig:s1schematic}(c) accordingly. Peak ZC (ZD) has a mixing frequency of $\omega_T=7.12$ ($\omega_T=-7.12$) THz, the frequency difference between the $5^2P_{1/2}$ and $5^2P_{3/2}$ states. 

The experimental 2D spectra can be reproduced in simulation based on the theoretical approach described in Section \ref{sec2}. The contribution from each pathway in Fig. \ref{fig:s1schematic}(c) needs to be calculated. As an example, assuming the excitation pulses are Delta pulses, the signal from the fourth diagram (SD2 shown in Fig. \ref{fig:s1schematic}-b) can be calculated by using the integrals in Eqs. (\ref{eq:integral1}-\ref{eq:integral4}) as:

\begin{widetext}
\begin{equation}\label{eq:7}
    \rho^{(4)}_{e_1e_1}(\tau,T,t)= S_0\mu_{e_1g}^{2}\mu_{ e_2 g}^{2} e^{-\Gamma_{e_1 e_1}\tau^{'}} e^{(-i\omega_{e_1 g} - \Gamma_{e_1 g})t} e^{(-i\omega_{e_1 e_2} - \Gamma_{e_1 e_2})T} e^{(-i\omega_{ g e_2} - \Gamma_{g e_2})\tau} \Theta(\tau^{'})\Theta(t)\Theta(T)\Theta(\tau),
\end{equation}
where
\begin{equation}
S_0 = -\frac{E_{A}E_{B}E_{C}E_{D}}{16\hbar^{4}}\rho_{gg}^{(0)} e^{-i(\mathbf{k}_{D}-\mathbf{k}_{C}-\mathbf{k}_{B}+\mathbf{k}_{A})\cdot \mathbf{r}}.
\end{equation}
\end{widetext}
Here $\tau^\prime$ is the fluorescence emission time, $E_i$ ($i=
A, B, C, D$) is the electric field amplitude for each pulse, $\rho_{gg}^{(0)}$ is the initial population in the ground state, $\mathbf{k}_i$ is the wave vector, and $\Theta$ is Heaviside step function. Taking the 2D Fourier Transform (2DFT) of Eq. (\ref{eq:7}) gives the fourth-order frequency-domain signal with $\tau^\prime=0$ as the following:

\begin{widetext}
\begin{eqnarray}
 S^{(4)}_{e_1e_1}(\omega_\tau,\omega_T,\omega_t)&=&S_0\mu_{e_1g}^{2}\mu_{ e_2 g}^{2}\bigg[\frac{1}{\omega_t-\omega_{e_1g}+i\Gamma_{e_1g}} \times\frac{1}{\omega_T-\omega_{e_1e_2}+i\Gamma_{e_1e_2}}\times\frac{1}{\omega_\tau-\omega_{g e_2 }+i\Gamma_{g e_2}}\bigg]
\end{eqnarray}
\end{widetext}

Repeating the above derivation for each pathway shown in Fig. \ref{fig:s1schematic}(c) and summing each contribution gives the overall nonlinear signal in the frequency domain. Evaluating the 2DFT of the time-domain expression at $T=0$ gives the single-quantum 2D frequency-domain spectrum solution shown in Eq. (\ref{eq:S1Signal}), while evaluating at $\tau=0$ gives the zero-quantum 2D frequency-domain solution shown in Eq. (\ref{eq:S0Signal}). Simulations for each spectra are shown in Fig. \ref{fig:s1spectra}(c) and Fig. \ref{fig:s0spectra}(b), respectively. 

\begin{eqnarray}
S^{(4)}_{1Q}(\omega_\tau,\omega_t)&=&\frac{2S_0\mu_{e_1g}^{4}}{\omega_t-\omega_{e_1g}+i\Gamma_{e_1g}} \times \frac{1}{\omega_\tau-\omega_{g e_1 }+i\Gamma_{g e_1}} \nonumber \\ &+&\frac{2S_0\mu_{e_1g}^{2}\mu_{ e_2 g}^{2}}{\omega_t-\omega_{e_1g}+i\Gamma_{e_1g}}\times \frac{1}{\omega_\tau-\omega_{g e_2 }+i\Gamma_{g e_2}}\nonumber \\&+&\frac{2S_0\mu_{e_1g}^{2}\mu_{ e_2 g}^{2}}{\omega_t-\omega_{e_2g}+i\Gamma_{e_2g}}\times \frac{1}{\omega_\tau-\omega_{g e_1 }+i\Gamma_{g e_1}}\nonumber \\&+&\frac{2S_0\mu_{ e_2 g}^{4}}{\omega_t-\omega_{e_2g}+i\Gamma_{e_2g}}\times \frac{1}{\omega_\tau-\omega_{g e_2 }+i\Gamma_{g e_2}}\nonumber  \\ \nonumber \\ \label{eq:S1Signal}
\end{eqnarray}
\begin{eqnarray}
S^{(4)}_{0Q}(\omega_T,\omega_t)&=&\frac{\mu_{e_1g}^{4}+\mu_{e_1g}^{2}\mu_{e_2g}^{2}}{\omega_t-\omega_{e_1g}+i\Gamma_{e_1g}}\times \frac{S_0}{\omega_T-\omega_{gg}+i\Gamma_{e_1e_2}} \nonumber \\&+& \frac{\mu_{e_1g}^{4}}{\omega_t-\omega_{e_1g}+i\Gamma_{e_1g}}\times \frac{S_0}{\omega_T-\omega_{e_1e_1}+i\Gamma_{e_1e_2}}
 \nonumber \\
 &+&  \frac{\mu_{e_1g}^{2}\mu_{ e_2 g}^{2}}{\omega_t-\omega_{e_1g}+i\Gamma_{e_1g}}\times \frac{S_0}{\omega_T-\omega_{e_1e_2}+i\Gamma_{e_1e_2}}\nonumber \\
 &+&\frac{\mu_{e_1g}^{2}\mu_{ e_2 g}^{2}}{\omega_t-\omega_{e_2g}+i\Gamma_{e_2g}}\times \frac{S_0}{\omega_T-\omega_{e_2e_1}+i\Gamma_{e_2e_1}}\nonumber \\ 
 &+&\frac{\mu_{e_1g}^{4}+\mu_{e_1g}^{2}\mu_{e_2g}^{2}}{\omega_t-\omega_{e_2g}+i\Gamma_{e_2g}}\times \frac{S_0}{\omega_T-\omega_{gg}+i\Gamma_{e_1e_2}}\nonumber \\  &+&\frac{\mu_{e_1g}^{4}}{\omega_t-\omega_{e_2g}+i\Gamma_{e_2g}}\times \frac{S_0}{\omega_T-\omega_{e_2e_2}+i\Gamma_{e_1e_2}}   \nonumber \\ \nonumber \\ \label{eq:S0Signal}
\end{eqnarray}

To conduct the single-quantum and zero-quantum simulations, the center frequencies for each resonance and their associated relaxation linewidths were extracted from the experimental spectra by fitting Lorentzian curves. Emission center frequencies $\omega_{e_{1}g} = 2\pi$ $\times$ 377.210 THz and $\omega_{e_{2}g} = 2\pi \times 384.245$ THz were extracted for resonances SA and SB shown in Fig. \ref{fig:s1spectra}. The relaxation linewidths $\Gamma_{e_{1}g}$ = $2\pi$ $\times$ 0.2446 THz and $\Gamma_{e_{2}g} = 2\pi \times 0.1450$ THz were extracted by slicing along the diagonal of SA and SB respectively. Slicing along the diagonal through SC and SD extracted parameters $\Gamma_{e_{1}e_{2}}$ = 2$\pi$ $\times$ 0.1446 THz and $\Gamma_{e_{2}e_{1}}$ = 2$\pi$ $\times$ 0.0716 THz respectively. The zero-quantum simulated spectra utilized the same parameters as the single-quantum simulation. The transition dipole moments were $\mu_{e_{1}g} = 2.537 \times 10^{-29}$ C$\cdot$m and $\mu_{e_{2}g} = 3.584 \times 10^{-29}$ C$\cdot$m. 

\section{\label{sec4}Double-quantum 2D spectra}
\begin{figure*}[tbh]
	\centering
	\includegraphics[width=\textwidth]{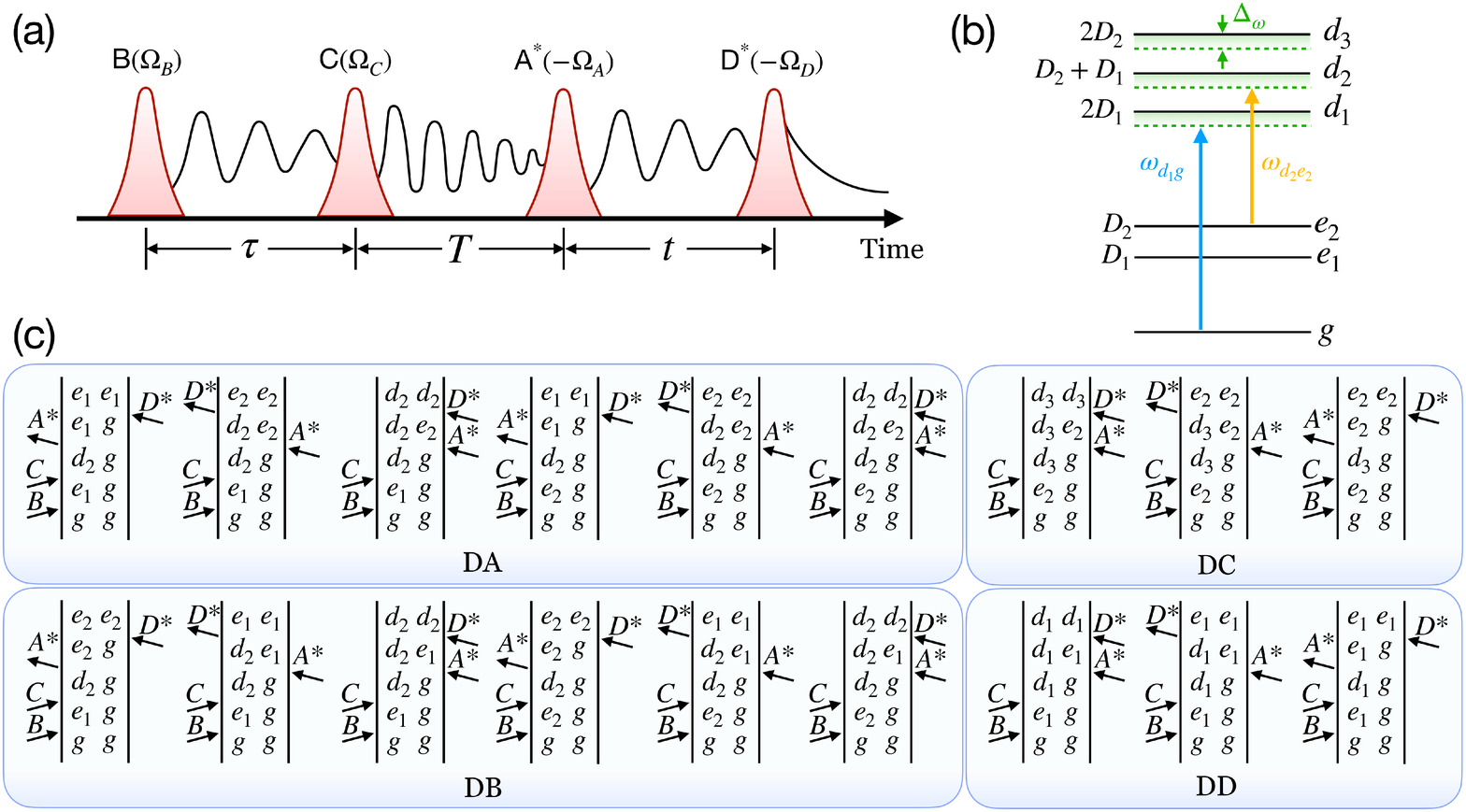}
	\caption{(a) Time ordering of excitation pulses for performing double-quantum 2DCS. (b) Relevant energy levels of Rb atoms including single-atom ($e_{1}$,$e_{2}$) and two-atom states ($d_{1}$,$d_{2}$,$d_{3}$). Two-atom interactions cause an energy shift ($\Delta_{\omega}$) to the two-atom states shown as the green dashed lines. No interactions would be described by the solid black lines. Shown are two example transitions between states g$\rightarrow d_{1}$ with energy $\omega_{d_{1}g}$ and another transition between states $e_{2} \rightarrow d_{2}$ with energy $\omega_{d_{2}e_{2}}$. (c) Double-sided Feynman diagrams representing possible excitation quantum pathways in double-quantum 2DCS using the pulse sequence in (a).}
	\label{fig:s3schematic}
\end{figure*}

Double-quantum 2D spectra were acquired by using the pulse sequence shown in Fig. \ref{fig:s3schematic}(a), where the conjugated pulses A and D arrive after pulses B and C. The first two pulses can excite  double-quantum coherence between the ground state and doubly excited states. The relevant energy levels, as shown in Fig. \ref{fig:s3schematic}(b), include the ground state $|g\rangle$, two singly excited states $|e_1\rangle=|5^2P_{1/2}\rangle$ and $|e_2\rangle=|5^2P_{3/2}\rangle$, and three doubly excited states which are two-atom states $|d_1\rangle=|5^2P_{1/2}, 5^2P_{1/2}\rangle$, $|d_2\rangle=|5^2P_{1/2}, 5^2P_{3/2}\rangle$, and $|d_3\rangle=|5^2P_{3/2}, 5^2P_{3/2}\rangle$. For convenience, the frequencies of $|e_1\rangle$ and $|e_2\rangle$ are labeled D$_1$ and D$_2$, respectively. The frequencies of the doubly excited states $|d_1\rangle$, $|d_2\rangle$, and $|d_3\rangle$ are 2D$_1$, D$_1+$D$_2$, and 2D$_2$, respectively. Using this pulse sequence, the excitation process is different from the single-quantum excitation. The first pulse, B, creates
single-quantum coherence between the ground state and the singly excited states. The second pulse, C, converts the single-quantum coherence to double-quantum coherence between the ground state and the doubly excited states. The third pulse, A, converts the double-quantum coherence back to single-quantum coherence. The fourth pulse, D, converts the single-quantum coherence to a population in the singly or doubly excited states, which emits a fluorescence signal. To detect the double-quantum signal, the fluorescence is measured by PD2 and its output is demodulated by a lock-in amplifier referenced to the mixing signal $\Omega_{S2}=\Omega_B+\Omega_C-\Omega_A-\Omega_D$. 
For the energy levels in Fig. \ref{fig:s3schematic}(b), this excitation process includes 18 pathways represented by the double-sided Feynman diagrams shown in Fig. \ref{fig:s3schematic}(c). In all pathways, a double-quantum coherence between the ground state and one of the doubly excited states evolves during the second time delay $T$. The dynamics of double-quantum coherence can be measured by scanning $T$ in double-quantum 2DCS. For two independent, non-interacting Rb atoms, the contributions from all 18 pathways cancel out, leading to a vanishing double-quantum signal. However, the cancellation is not complete if there is the interaction between the Rb atoms that breaks the symmetry, resulting in a non-zero double-quantum signal. Double-quantum 2DCS has been proven to be extremely sensitive detection to dipole-dipole interactions and collective resonances in dilute K and Rb atomic vapors \cite{Dai2012a,Gao:16,PhysRevLett.120.233401,Yu2018,Liang2021,PhysRevLett.128.103601}.  

\begin{figure}[H]
	\centering
	\includegraphics[width=\columnwidth]{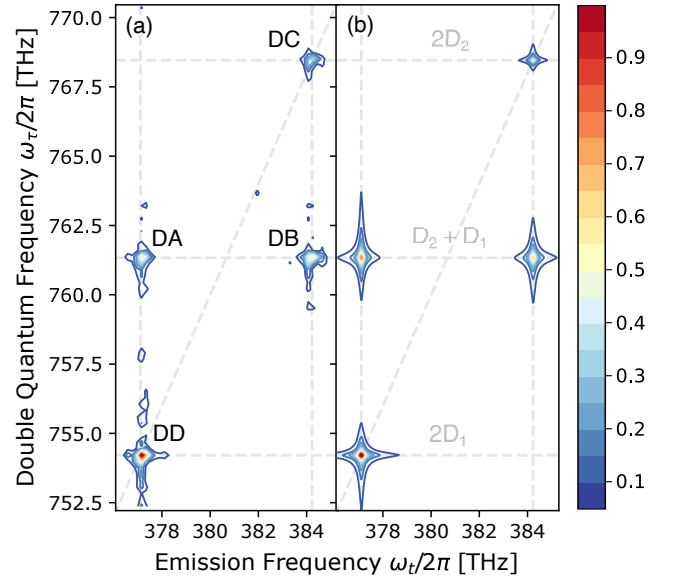}
	\caption{(a) Experimental and (b) simulated double-quantum 2D spectra of a Rb vapor. The spectral amplitude is plotted with the maximum normalized to 1.}
	\label{fig:s3spectra}
\end{figure}

The double-quantum signal was measured as time delays $T$ and $t$ are scanned while time delay $\tau$ is fixed. Fourier-transforming the signal into the frequency domain generates double-quantum 2D spectra. A typical double-quantum 2D spectrum is shown in Fig. \ref{fig:s3spectra}(a), where the spectral amplitude is plotted with the maximum normalized to 1. The $x$ and $y$ axes are the emission frequency $\omega_t$ and the double-quantum frequency $\omega_T$, corresponding to the time delays $t$ and $T$. There are four peaks in the double-quantum 2D spectrum. The double-quantum frequencies of these peaks match the frequencies of the two-atom doubly excited states 2D$_1$, D$_1+$D$_2$, and 2D$_2$, since the double-quantum coherence oscillates at these frequencies during time delay $T$. The emission frequencies of these peaks are D$_1$ and D$_2$. Each peak has contributions from multiple pathways represented by double-sided Feynman diagrams labeled accordingly in Fig. \ref{fig:s3schematic}(c). The double-quantum 2D spectrum reveals the two-atom collective resonances and dipole-dipole interactions involving both $5^2P_{1/2}$ and $5^2P_{3/2}$ states. For Rb atoms, there is also a single-atom doubly excited state $5^2D$ at $2\pi \times 770.5$ THz. The double-quantum signal associated with $5^2D$ has been observed before \cite{Gao:16} as two off-diagonal peaks with the double-quantum frequency at $\omega_T=2\pi \times 770.5$ THz and the emission frequency at $\omega_t= 2\pi \times 384.2$ and $2\pi \times 386.3$ THz. However, in the current experiment, the laser spectrum was centered at $\sim$810 nm and does not have sufficient intensity at the wavelength required for exciting $5^2D$. The double-quantum signal associated with $5^2D$ was not observed here. 

The experimental frequency-domain double-quantum 2D spectrum can be reproduced by calculating contributions from all pathways in Fig. \ref{fig:s3schematic}(c). 
\begin{eqnarray}
&S^{(4)}_{DA}(\omega_T,\omega_t)&=\frac{S_0(\mu_{e_1 g}\mu_{d_2 e_1}+\mu_{e_2 g}\mu_{d_2 e_2})}{\omega_T-\omega_{d_2 g }+i\Gamma_{d_2 g}} \\
&&\times \bigg(\frac{\mu_{e_2 g}\mu_{d_2e_2}}{\omega_t-\omega_{d_2 e_2 }+i\Gamma_{d_2 e_2}}- \frac{ \mu_{d_2 e_1}\mu_{ge_1}}{\omega_t-\omega_{e_1 g }+i\Gamma_{e_1 g}}\bigg) \nonumber \\
&S^{(4)}_{DB}(\omega_T,\omega_t)&=\frac{S_0(\mu_{e_1 g}\mu_{d_2 e_1}+\mu_{e_2 g}\mu_{d_2 e_2})}{\omega_T-\omega_{d_2 g }+i\Gamma_{d_2 g}} \\
&&\times \bigg(\frac{\mu_{e_1 g}\mu_{d_2e_1}}{\omega_t-\omega_{d_2 e_1 }+i\Gamma_{d_2 e_1}}-\frac{ \mu_{d_2 e_2}\mu_{e_2g}}{\omega_t-\omega_{e_2 g }+i\Gamma_{e_2 g}}\bigg) \nonumber \\
&S^{(4)}_{DC}(\omega_T,\omega_t)&=\frac{S_0\mu^2_{e_2 g}\mu^2_{d_3 e_2}}{\omega_T-\omega_{d_3 g }+i\Gamma_{d_3 g}} \\
&&\times \bigg(\frac{1}{\omega_t-\omega_{d_3 e_2 }+i\Gamma_{d_3 e_2}}-\frac{1}{\omega_t-\omega_{e_2 g }+i\Gamma_{e_2 g}}  \bigg)\nonumber \\
&S^{(4)}_{DD}(\omega_T,\omega_t)&=\frac{S_0\mu^2_{e_1 g}\mu^2_{d_1 e_1}}{\omega_T-\omega_{d_1 g }+i\Gamma_{d_1 g}} \\
&&\times \bigg( \frac{1}{\omega_t-\omega_{d_1 e_1 }+i\Gamma_{d_1 e_1}}-\frac{1}{\omega_t-\omega_{e_1 g }+i\Gamma_{e_1 g}}  \bigg)\nonumber
\end{eqnarray}

The simulated double-quantum spectra used parameters extracted from the experimental spectra with fitted Lorentzian curves. The double-quantum center frequencies extracted from the experimental spectra were $\omega_{d_{1}g} = 2\pi \times$ 754.213 THz, $\omega_{d_{2}g} = 2\pi \times$ 761.358 THz, and $\omega_{d_{3}g} = 2\pi \times$ 768.509 THz. The emission center frequencies ($\omega_{e_{1}g}$, $\omega_{e_{2}g}$) were identical to those extracted from the single-quantum spectra. The remaining frequency terms are 
\begin{eqnarray}
\omega_{d_{3}e_{2}} &= \omega_{e_{2}g} - \Delta_{\omega} \\ \omega_{d_{2}e_{2}} &= \omega_{e_{1}g} - \Delta_{\omega} \\
\omega_{d_{2}e_{1}} &= \omega_{e_{2}g} - \Delta_{\omega} \\ \omega_{d_{1}e_{1}} &= \omega_{e_{1}g} - \Delta_{\omega}
\end{eqnarray}
where $\Delta_{\omega} = 2\pi \times 100$ MHz. The terms $\omega_{e_{1}g}$ and $\omega_{e_{2}g}$ are the same used for the single-quantum spectra simulation. The remaining relaxation terms are 
\begin{eqnarray}
\Gamma_{d_{3}e_{2}} = \Gamma_{e_{2}g} + \Delta_{\Gamma} \\ \Gamma_{d_{2}e_{2}} = \Gamma_{e_{1}g} + \Delta_{\Gamma} \\ \Gamma_{d_{2}e_{1}} = \Gamma_{e_{2}g} + \Delta_{\Gamma} \\ \Gamma_{d_{1}e_{1}} = \Gamma_{e_{1}g} + \Delta_{\Gamma}
\end{eqnarray}
where $\Delta_{\Gamma} = 2\pi \times 15$ GHz. The terms $\Gamma_{e_{1}g}$ and $\Gamma_{e_{2}g}$ were the values extracted from the experimental single-quantum spectra. 

The energy splitting $\Delta_{\omega}$ and relaxation shift $\Delta_{\Gamma}$ are brought about by the interaction of individual atoms. A case where $\Delta_{\omega}$ = $\Delta_{\Gamma}$ = 0 would suggest no interactions between individual atoms. This would lead to a double-quantum spectra with S$^{(4)}_{DD}$ = S$^{(4)}_{DC}$ = 0 and S$^{(4)}_{DA}$ = S$^{(4)}_{DB} \neq$  0 which is not reflected by the spectra in Fig. \ref{fig:s3spectra}. 

\section{\label{sec5}Conclusion}
In conclusion, we have implemented a broadband collinear optical 2DCS experiment on Rb atoms and obtained a complete set of single-quantum, zero-quantum, and double-quantum 2D spectra including both D-line transitions of Rb. The single-quantum 2D spectrum shows the coherent coupling between two D-line transitions. The zero-quantum 2D spectrum reveals the coherence between two excited states. The double-quantum 2D spectrum is a result of the dipole-dipole interaction and collective resonance between two atoms. Simulated 2D spectra based on the perturbative solutions to the OBEs agree well with the experimental spectra. The measurements in Rb atoms complement previous 2DCS studies of K and Rb with a narrower bandwidth that covers two D-lines of K or only a single D-line of Rb. The broadband excitation enables the possibility of double-quantum and multi-quantum 2DCS of both D-lines of Rb to study many-body interactions and correlations in comparison with K atoms.

\begin{acknowledgments}
This work was partially supported by NSF under Grants PHY-1707364 and CHE-2003785.
\end{acknowledgments}

%\appendix

%\section{Appendixes}

%\bibliography{rb2dcs}% Produces the bibliography via BibTeX.

%

\end{document}